# Anomalous electronic susceptibility in $Bi_2Sr_2CuO_{6+\delta}$ and comparison with other overdoped cuprates

G. Le Bras, Z. Konstantinovic, D. Colson, A. Forget, J-P. Carton, and C. Ayache

Service de Physique de l'Etat Condensé, CEA Saclay ; 91191 Gif sur Yvette, France

F. Jean[1,2], G. Collin[2]

LEMHE, CNRS UMR 8647, Bât 415, Université Paris Sud, 91405 Orsay, France
LLB, CEA-CNRS, CE Saclay ; 91191 Gif sur Yvette, France

Y. Dumont

LMOV, Université de Versailles-Saint Quentin en Yvelines, 45 Av. des Etats Unis, 78035 Versailles, France

We report magnetic susceptibility performed on overdoped $Bi_2Sr_2CuO_{6+\delta}$ powders as a function of oxygen doping $\delta$ and temperature T. The decrease of the spin susceptibility $\chi_s$ with increasing T is confirmed. At sufficient high temperature, $\chi_s$ presents an unusual linear temperature dependence $\chi_s \sim \chi_{s0} - \chi_1 T$. Moreover, a linear correlation between $\chi_1$ and $\chi_{s0}$ for increasing hole concentration is displayed. A temperature $T_? = -(d\chi_{s0}/d\chi_1)$, independent of hole doping characterizes this scaling. Comparison with other cuprates of the literature ($La_{2-x}Sr_xCuO_4$, $Tl_2Ba_2CuO_{6+\delta}$ and $Bi_2Sr_2CaCu_2O_{8+\delta}$), over the same overdoped range, shows similarities with above results. These non conventional metal features will be discussed in terms of a singular narrow-band structure.

It is widely believed that the study of the unusual normal state properties in cuprates is very important to understand the mechanism that underlies high-$T_c$ superconductivity. The underdoped part of the generic critical temperature carrier density phase diagram of high-$T_c$ cuprate superconductors is characterized by the pseudogap effects which has been extensively investigated through various experimental approaches. The overdoped regime, considered as having the "classical" electronic properties of a normal metal, has been much less studied.

Recent theoretical and experimental results have shown that there is considerable evidence to the contrary. In a conventional Fermi Liquid (FL), both electronic spin susceptibility ($\chi_s$) and electronic specific heat coefficient ($\gamma$) are, in a first approximation, independent of T. In the cuprates, by contrast, $\chi_s$ and $\gamma$ present both positive slopes in the underdoped regime[1] and negative slopes in the overdoped one[2,3,4,5]. At optimal hole concentration, where $T_c$ takes on its maximum value, $\chi_s$ and $\gamma$ remain temperature independent. Moreover, the first temperature corrections for several physical parameters measured in the overdoped range do not correspond to those of an usual FL. For instance, the electronic spin susceptibility, $\chi_s$, shows a close to linear decrease with T in this overdoped range[6]. This trend can be also related to a similar behavior observed in the electronic specific heat and in the Knight shift[7,8] as well as to the anomalous T-dependence of the thermopower in this doping range[9,10].

On the other hand, the overall $\chi_s$[5] and $\gamma$[11] magnitudes increase with the oxygen content, $\delta$, even beyond the hole doping corresponding to $T_c^{max}$. In the case of $La_{2-x}Sr_xCuO_4$ (LSCO), both $\chi_s$ and $\gamma$ then reach a maximum and decrease at higher doping[11,12,13]. These untypical hole doping and temperature dependencies suggest a departure from a conventional FL, not only for underdoped range, but also in the overdoped one. A strong increase in the electronic density of states (DOS) with doping has been proposed as a possible origin for these unconventional properties[11]. Such an hypothesis can be fruitfully investigated by comparison with recent progress obtained through angle-resolved photoemission (ARPES) measurements in this overdoped range.

In this paper, we report measurements of bulk magnetic susceptibility as a function of temperature and doping in the normal state of overdoped $Bi_2Sr_2CuO_{6+\delta}$ (Bi-2201) powders. We have shown that the electronic susceptibility as a function of T and p presents a unconventional Fermi liquid behavior arising from electronic effects. The comparaison of our results with those presented in the literature for other cuprates reveals an universal trends of the susceptibility in the overdoped cuprates.

The compound Bi-2201 is a convenient system for studying the normal state in the overdoped regime. Indeed, in its as grown form (i.e. non substitued), this compound lies naturally in the overdoped range. It has a low critical temperature ($T_{cmax} \sim 20K$), allowing the study of normal state properties over an extended temperature range. Polycrystalline samples were prepared by the classical solid-state reaction method [14].

The absolute content of oxygen nonstoichiometry $\delta$ of the compound $Bi_2Sr_2CuO_{6+\delta}$ have been investigated by thermogravimetric techniques from 300°C to 700°C with oxygen partial pressure $P_{O2}$ ranging from $10^{-4}$ atm to 1 atm under thermodynamic equilibrium conditions. We have explored the T-$\delta$ phase diagram for oxygen concentration varying from $\delta=0.09$ to 0.18. Since the Bi-2201 compound does not contain CuO chains, we can directly study the influence of the hole doping on the properties of the $CuO_2$ planes.

Samples were characterized by powder X-ray diffraction revealing that all the peaks can be attributed to a Bi-2201 single phase[15]. The critical temperature is determined by the onset in the Zero Field Cooled (ZFC) susceptibility measurements. The hole concentration in the $CuO_2$ planes has been estimated by thermoelectric power measurements performed on sintered samples made from the same powder and submitted to the same oxygen treatment[10]. We found that p changes from 0.20 to 0.29 causing a decrease from $T_c \sim 16K$ to $T_c<1.5K$.

Susceptibility measurements in the metallic state were performed using a commercial Squid magnetometer (Cryogenics). Figure 1 shows the temperature dependence between 50 K and 300 K of the normal-state susceptibility under 5T for the various hole concentrations. As Lee, Klemm and Johnston[16] have shown that the normal state magnetic susceptibility of the superconducting oxides is altered for temperatures below $\sim 2T_c$ by the diamagnetism due to superconducting fluctuations, results below 50 K are omitted. A sizable Curie-Weiss contribution (~ 0.5%) which presumably arises from paramagnetic impurities or disorder effects[2] has been substracted from the susceptibility data shown in figure 1.

$\chi(T)$ in the close-to-optimal doping sample (p ~ 0.20) has moderate temperature dependence with a positive slope in the range of temperature 50 K-200 K which may be a reminiscence of the pseudogap effect. On increasing hole concentration, $\chi(T)$ increases with decreasing temperature. The negative slope $d\chi/dT$ observed at room temperature (RT) is a universal trend observed in all overdoped cuprates. The susceptibility, corrected from the Curie-Weiss contribution, can be written, in a first approximation:

$\chi = \chi_{core} + \chi_{vv} + \chi_{spin}$ (1)

where $\chi_{core}$, $\chi_{vv}$ and $\chi_{spin}$ are respectively the core diamagnetism, the Van Vleck paramagnetism and the spin terms. Since the Van Vleck paramagnetism and the core diamagnetism are essentially temperature independent (the values of these contributions are listed in the table 1 of Ref.2) the temperature dependence of the susceptibility arises from the spin contribution and can be linearly fitted :

$\chi_{spin}(T) \sim \chi_{s0} - \chi_1 T$ (2)

The linear temperature dependence $\chi_1 T$ is unusual and was first found by Allgeier *et al*[2]. It does not correspond to the first temperature correction of the susceptibility in $T^2$, as observed in a conventional metal.

Parameters $\chi_{s0}$ (Fig. 2a), $\chi_1$ (Fig. 2b) and $T_c^{max}$ (insert in Fig. 2b) are reported as a function of p for Bi-2201 (our data) and for LSCO compounds (data from Ref. 11) treated in the same way. In the case of LSCO, Torrance[5] as well as Loram[11] have reported the hole doping dependence of the electronic susceptibility for the entire generic critical temperature carrier density phase diagram and have shown that there is no significant change at $T_{cmax}$. The present work is focused on the "overdoped regime" ie p>0.16, and the hole doping dependence of the two contributions to the spin susceptibility are shown separately.

The first thing to note is that $\chi_{s0}$ as well as $\chi_1$ of both compounds have a similar behavior as a function of p. At p~0.16, $T_c$ is maximum and $\chi_1$ is nearly 0 (as shown above in Fig.1, $\chi$ is nearly temperature independent at optimal doping).

For the Bi-2201 compound, $\chi_{s0}$ and $\chi_1$ increase drastically for all hole concentration whereas in the case of LSCO, $\chi_{s0}$ and $\chi_1$ reach a plateau at p~0.22 which is extended until p~0.32; the middle of the plateau corresponds to the characteristic hole doping p~0.27. At p~0.32, $\chi_{s0}$ and $\chi_1$ begin to decrease; $\chi_1$ falls to 0 at p~0.45 which means that the electronic susceptibility recovers a classical Fermi-liquid behavior, independent of the temperature only well above the vanishing of the superconductivity.

Ino *et al*[12,17] have shown the same type of hole doping dependence in the spin susceptibility of LSCO. The plateau is not clearly present in the spin susceptibility *vs.* p as reported by Torrance *et al*[5], however they observed a more smooth increase of $\chi_s$ for 0.22<p <0.25 than at lower doping and a decrease for p≥ 0.27. It seems clear therefore that the hole doping dependence of the overdoped Bi-2201 and LSCO compounds are rather different.

Specific heat measurements in the overdoped compound LSCO have displayed a similar hole doping for $\gamma$ and for $\chi_s$[11,13] which indicates the same electronic origin. The magnitude of the electronic specific heat coefficient $\gamma$ at T=0K and the spin susceptibility are related by a factor close to the free electron Wilson ratio[11] ($a_o = \pi^2/3 \, (k_B/\mu_B)^2$) $\gamma_0 = a_0 \chi_{s0}$. We have therefore extracted $\gamma_0$ of the Bi-2201 compound from our data of susceptibility for different hole doping. The achieved values (reported on the right scale of the figure 2b) are of the same magnitude as those directly obtained by specific heat measurements[18] (open circle in Fig.2a). This result shows that the values of $\chi_{s0}$ obtained by fitting susceptibility in the range 50 K-RT with the unusual T-linear term gives rise to the correct order of magnitude for $\gamma_0$.

In figure 3, we have reported $\chi_1$ versus $\chi_{s0}$ for Bi-2201 (solid circle), and LSC0 (open triangle) for different oxygen contents. The data of susceptiblities for the overdoped $Tl_2Ba_2CuO_{6+\delta}$ (Tl-2201) powders[19] (0.20<p≤0.30) and $Bi_2Sr_2CaCu_2O_{8+\delta}$ (Bi-2212) single crystals[20] (0.18≤p≤0.22), that can be fitted with a linear temperature term, have been analysed in the same way as shown in insert of Fig. 3.

Remarkably, below p~0.3, all compounds show a linear increase of $\chi_1$ with $\chi_{s0}$ (and increasing hole concentration). In the case of LSCO, for which data above p=0.3 exists, a turning point is observed around this value, which also corresponds to the vanishing of superconductivity. For higher doping (p ≥ 0.3), the magnitude of $\chi_1$ and $\chi_{s0}$ decreases ( as also shown in Fig.2a and 2b) until $\chi_1$ falls to 0 at p=0.45.

This linear correlation between $\chi_1$ and $\chi_{s0}$ on a large range of hole doping seems therefore to be a general trend in the overdoped cuprates. It means that the hole doping dependence of $\chi_1$ shown in figure 2a arises in fact from the hole doping dependence of $\chi_{s0}$. This result, connected to the hole doping variation of the electronic specific heat coefficient, confirm that the two terms of $\chi_s(T)$ originate from electronic contributions. In the range of hole doping where $\chi_1$ linearly depends of $\chi_{s0}$, we have defined a characteristic temperature $T_\chi = (d\chi_{s0}/d\chi_1)$. By order of its definition, $T_\chi$ is hole doping independent. The extrapolated linear fits all meet at $T_\chi$:

$\chi_{spin}(T) \sim \chi_1 (T_\chi - T) + $ Constant  (3)

We have determined $T_\chi$ for the four overdoped cuprates mentioned before, namely the Bi-2201, LSCO, Tl-22101 powders and the Bi-2212 single crystal (in this last case, $T_\chi$ corresponds to $T_\chi^{ab}$). Figure 4 shows an apparent correlation between $T_\chi$ and $T_c^{max}$. $T_\chi$ increases with increasing maximal critical temperature. On the other hand, the value of $T_\chi$ seems not to depend of the number of $CuO_2$ planes in a unit cell. Indeed, the value of $T_\chi$ obtained for Tl-2201 is more than twice as high that the values obtained for Bi-2201 and LSCO compounds containing, as well as Tl-2201, one $CuO_2$ plane in a unit cell.

$\chi_s$ reflects the density of states averaged over an energy range ~ ($E_F \pm 2k_BT$). Thus the behavior of $\chi_s$ as a function of hole doping (Fig.2) should reflect the presence of a peak in the energy dependence of the DOS. Such a deduction is in good agreement with ARPES measurements on HTSC that have allowed to identify the presence of saddle points centered at k= ($\pi$, 0) and (0, $\pi$) in the band structure which give rise to van Hove singularities (VHS) in the DOS near the Fermi level $E_F$. In the case of Bi-2201, a temperature dependent ARPES study, with various doping from underdoped state to strongly overdoped state, has shown that this VHS lies below $E_F$; this means that the Fermi Surface (FS) retains its hole like character centered at ($\pi,\pi$) for all doping levels [21, 22] and, consequently, $\chi_s$ increases continuously as a function of p. Temperature dependent ARPES measurements have shown that in the overdoped superstructure-free $Bi_{1.8}Pb_{0.38}Sr_{2.01}CuO_{6-\delta}$ ($T_c$<4K), the bottom of the dispersive band at ($\pi$,0) is located at about 5 meV below $E_F$ [22]. This result suggests that a more heavily overdoped sample of Bi-2201 should have an electron like FS and a decrease of $\chi_s$ is therefore expected at higher doping.

On the other hand, in LSCO, it has been unambiguously observed that a band of flat dispersion at ($\pi$, 0) moves from below $E_F$ for p<0.2 to above $E_F$ for p>0.2, so that the topological center of the FS is turned over from ($\pi,\pi$) to (0,0) and the hole-like FS becomes electron-like [23,24]. This change of character of the FS gives rise to a decrease of the susceptibility (as well as to a decrease of the electronic specific heat[11,13]), however, not as expected, at the characteristic hole doping p~0.2 corresponding at the point where the VHS and the Fermi level coincides, but for p>0.3. Indeed, at p~0.2, $\chi_s$ reachs a plateau extended from 0.2<p<0.3. This plateau might results from the three-dimensionnal character of LSCO over the whole temperature range in the overdoped regime as displayed by anisotropic transport measurements [25]. Moreover, it should be noticed that this change of topology of FS at p~0.2 is accompanied of an orthorombic to tetragonal structural phase transition (SPT)[26] which may lead to other complicated effects.

It therefore seems on one hand, that the behavior of $\chi_s$ as a function of hole doping, is strongly related to the topology of the Fermi Surface. On the other hand, we have shown that the change of the topology of FS appears well above $p(T_{cmax})$ in the case of LSCO and certainly at more higher doping for Bi-2201. This result is surprising as it is in contradiction with several theoritical predictions which indicate that the maximum of $T_c$ corresponds to a maximum in the DOS[27,28].

In a metal with a constant density of states near the Fermi level, the electronic susceptibility $\chi_s$ is temperature independent in contradiction with observations; we have shown above a linear decreasing of $\chi_s$ with increasing temperature in the overdoped cuprates. This unusual linear temperature dependence has also been provided by a NMR study[7] (that shown that the Knight shift and therefore the spin susceptibility can be expressed in the same way). Specific heat measurements in LSCO[11] display the same linear temperature dependence of the electronic specific heat coefficient. Recent thermoelectric power measurements on the same Bi-2201 samples as those presented in this paper, reveals a linear variation of the diffusion contribution $T/S_{diff}$. It therefore seems that there exists in the overdoped cuprates a general behavior of the normal state properties with a linear T dependence.

J. Bouvier and J. Bok[29] have used the van Hove theory to investigate the Pauli spin susceptibility as a function of the temperature for different doping in HTSC. In this scenario which used a rigid band model, the effect of overdoping is to displace the Fermi level $E_F$ from the singularity $E_s$. The VHS has been fixed at $E_F$ for p=0.20 hole/copper atom in the $CuO_2$ plane. In this context, they found that in the overdoped regime p<0.2, and at sufficient high temperature, $\chi_s$ increases with decreasing temperature (Fig.1 of the ref. 29), in good agreement with our experimental observations. However, in this range of temperature, for which $\chi_s$ increases with decreasing T, the curves have been calculated with a law in $\ln(1/T)$ (resulting from the VHS logarithmic DOS) which can not be used to fit our experimental susceptibility curves. This suggests that the singularity in the DOS might be of an another type. The difference could be ascribed to many-body effects, such as quasiparticle decay which, morover, are modified in the vicinity of a VHS. But, our formula (3) reveals an intriguing scaling in which doping and temperature dependences are well separated, and that whatever explanation should account for.

In conclusion, the present study features that in the overdoped regime, the spin susceptibility as a function of doping p and temperature T shows a departure from the conventional FL which can be interpreted by the presence of a growing peak near $E_F$ in the DOS, in good agreement with the ARPES observations. We have also shown that the maximum of this peak is not located at $p(T_{cmax})$ but seems to be related to the change of the topology of the Fermi surface. A VHS in a rigid band model does not allow to fit our experimental curves. More complicated effects such as the coulomb interaction or the life-time of the quasi-particles should certainly be taken in account. It is important to note that we have defined in numerous overdoped systems a characteristic temperature independent of the hole concentration, that might be correlated to $T_{cmax}$.

The authors thank J. Bok, J. Bouvier, M. Norman, C. Pépin and M. Roger for fruitful discussions, P. Monod , for RPE measurements on the Bi-2201 samples and L. Le Pape for his technical support.

[1] See for a review, T. Timusk and B. Statt, Rep. Prog. Phys. **62**, 61 (1999)

[2] C. Allgeier, and J. S. Schilling, Phys. Rev. B **48,** 9747 (1993).

[3] W. C. Klee, R. A. Klemm, and D. C. Johnston , Phys. Rev. Lett. **63**, 1012 (1989).

[4] M. Oda, T. Nakano, Y. Kamada, and M. Ido, Physica C **183**, 234-240 (1991).

[5] J. B. Torrance *et al*., Phys. Rev. B **40**, 8872 (1989).

[6] I.R. Fisher, J. R. Cooper, R. J. Cava Phys. Rev. B**52**, 15086(1995).

[7] A. Trokiner *et al.,* Phys. Rev. B **41,** 9570 (1990).

[8] P. V. Bellot *et al*., Physica C **282-287**, 1357 (1997), thèse P. V Bellot (1997) ESPCI and Université Paris 6, France.

[9] M. Y. Choi and J. S. Kim, Phys. Rev. B **59**, 192 (1999).

[10] Z. Konstantinovic *et al*, Phys Rev. B., in press

[11] J. W. Loram, K. A. Mirza, J. R. Cooper, N. Athanassopoulou and W. Y. Liang, Proc. 10$^{th}$ Anniversary workshop on Physics, Materials and Application, ed. B. Battlog, C. W. Chu, W. K. Chu, D. U. Gubser and K. A. Müller (1996).

[12] T. Nakano *et al.* Phys. Rev. B **49**, 16000 (1994).

[13] N. Momono *et al.* Physica C **233** 395 (1994).

[14] F. Jean, G. Collin, M. Andrieux, N. Blanchard and J. F. Marucco, Physica C. **339** (4) , 269 (2000).

[15] F. Jean *et al. to be published*

[16] W. C. Lee, R. A. Klemm, D. C. Johnston, Phys. Rev. Lett. **63**, 1012 (1989).

[17] A. Ino, *et al.*. in Phys. Rev. Lett **81** 2124 (1998).

[18] Thèse E. Janod (1996) Laboratoire de Cryophysique, SPSMS, CEA Grenoble, France and Université Joseph Fourier-Grenoble I.

[19] Y. Kubo, Y. Shimakawa, T. Kondo, and H. Igarashi, Physica C **185-189**, 1253 (1991).

[20] T. Watanabe, T. Fudgi and A. Matsuda, Phys. Rev. Lett. **84**, 5848 (2000).

[21] T. Sato, T. Kamiyama, Y. Naitoh, and T. Takahashi, Phys. Rev. B. **63** 132502 (2001).

[22] T. Sato *et al*., Phys. Rev. B. **64** 054502-1 (2001).

[23] T. Yoshida *et al.*., Phys. Rev. B **63**, 220501 (R) (2001).

[24] A. Ino *et al.,* Phys. Rev. B, **65,** 094504 (2002).

[25] Y. Nakamura and S. Uchida, Phys. Rev. B **47**, 8369 (1993).

[26] J. Labbé and J. Bok, Europhys. Lett. **3**(11), 1225 (1987).

[27] D. M. Newns, P. C. Pattnaik, C. C. Tsuei, Phys. Rev. B. **43**, 3075 (1991).

[28] D. M. Newns, C. C. Tsuei, and P. C. Pattnaik, Phys. Rev. B. **52**, 13611 (1995).

[29] J. Bouvier, and J. Bok Journal of Superconductivity **10** 673 (1997).

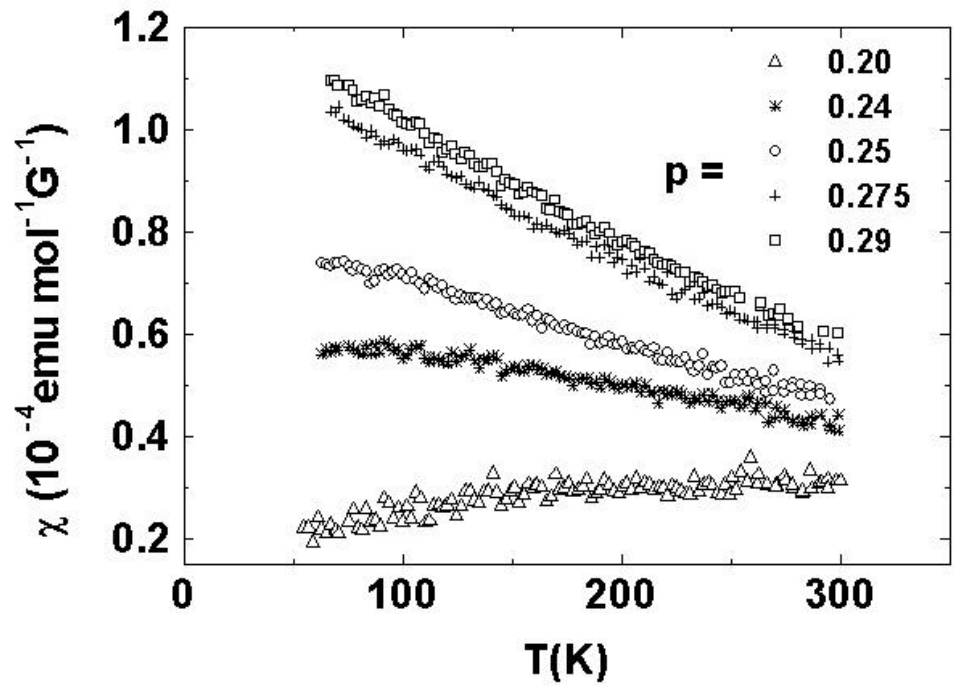

**Figure 1: Magnetic susceptibilities of the polycrystalline samples $Bi_2Sr_2CuO_{6+d}$ for various hole doping p under H=5T. The Curie-Weiss contribution has been substracted.**

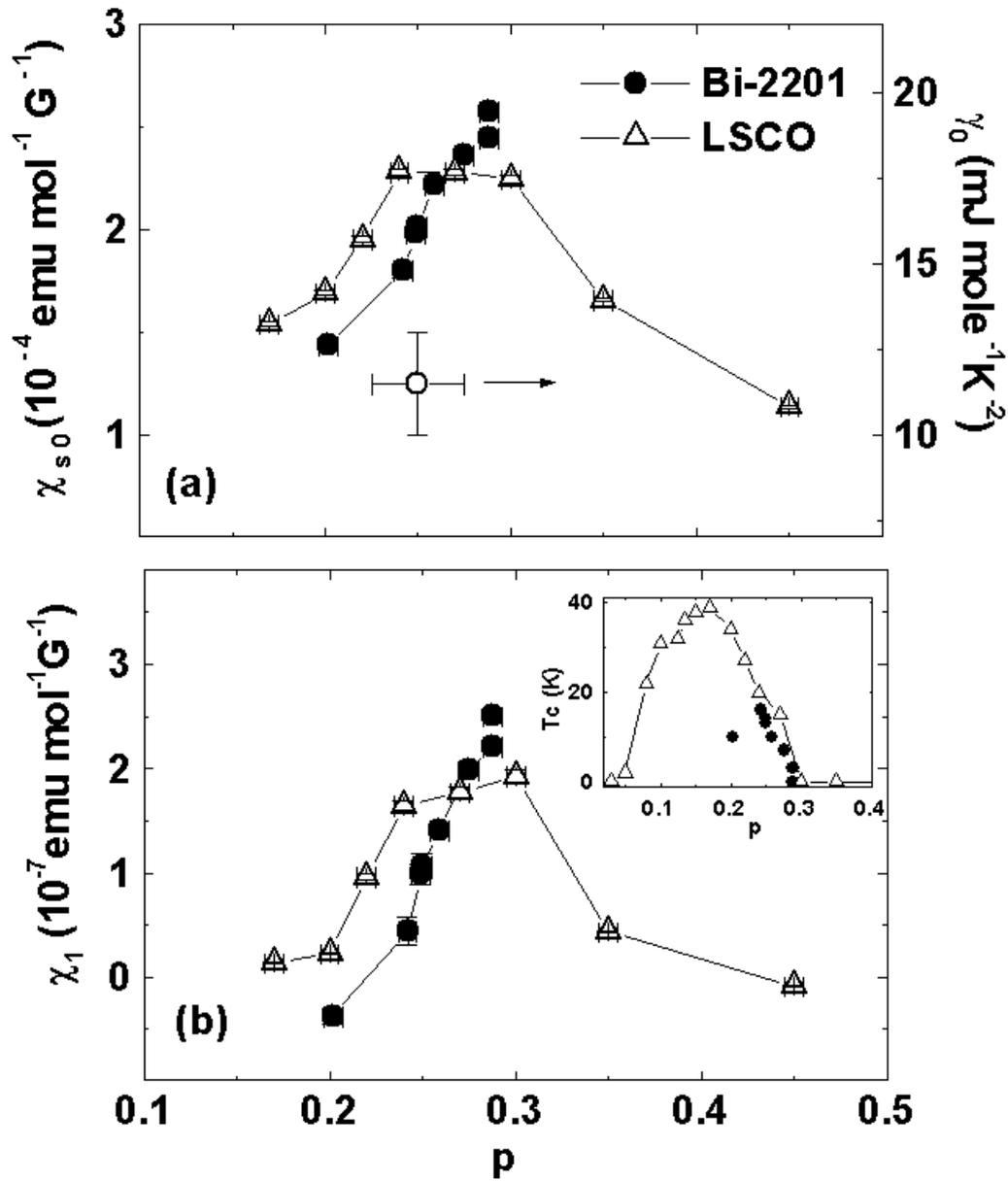

**Figure 2:** Hole doping dependence of the two contributions to the electronic susceptibility $\chi_{s0}$ (a) and $\chi_1$ (b) of Bi-2201(our data) and $La_{2-x}Sr_xCuO_4$ (Loram data). In a) the right scale gives the value of the electronic specific heat ratio $\gamma(T=0K)$ obtained via $\chi_{s0}$; the open circles correspond to the $\gamma(T=0K)$ of Bi-2201[18]. As an insert in (b), we have reported the hole doping dependence of Tc for both compounds.

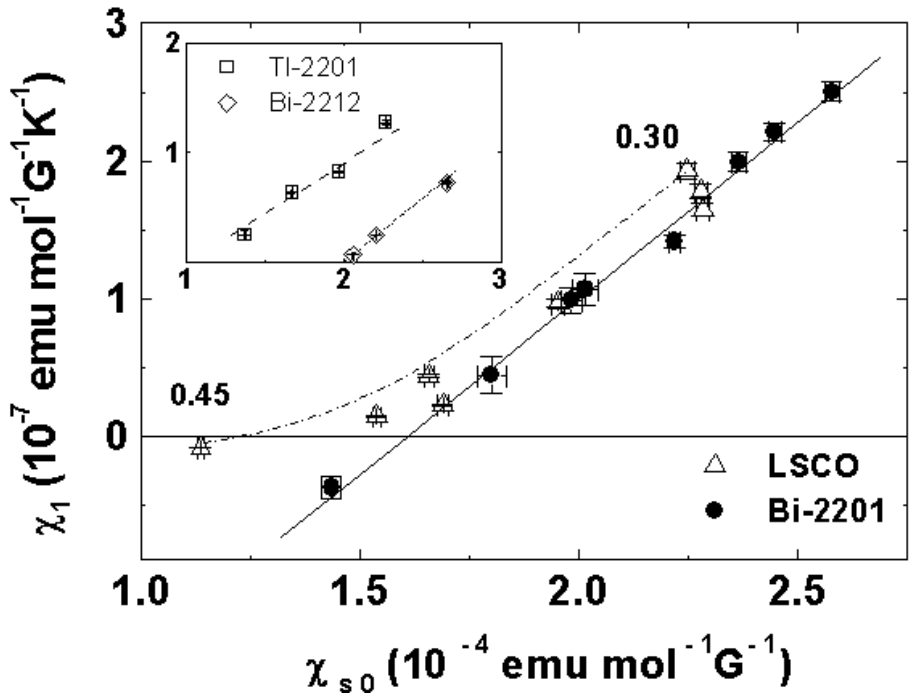

Figure 3: $c_1$ versus $c_{s0}$ for both overdoped compounds Bi-2201(our data) and LSCO[11]. The black line represents the linear fit in the range of hole doping 0.20<p<0.28 for Bi-2201. The dash dot line is only a guide for the eye that connects the points corresponding at the decrease of $c_{s0}$ (and $c_1$) when p>0.3. In the insert, we report $c_1$ versus $c_{s0}$ for both overdoped compounds Tl-2201[19] and Bi-2212[20]. The labels of the scale are the same as those of the main figure. The dotted and dashed lines correspond to a linear fit.

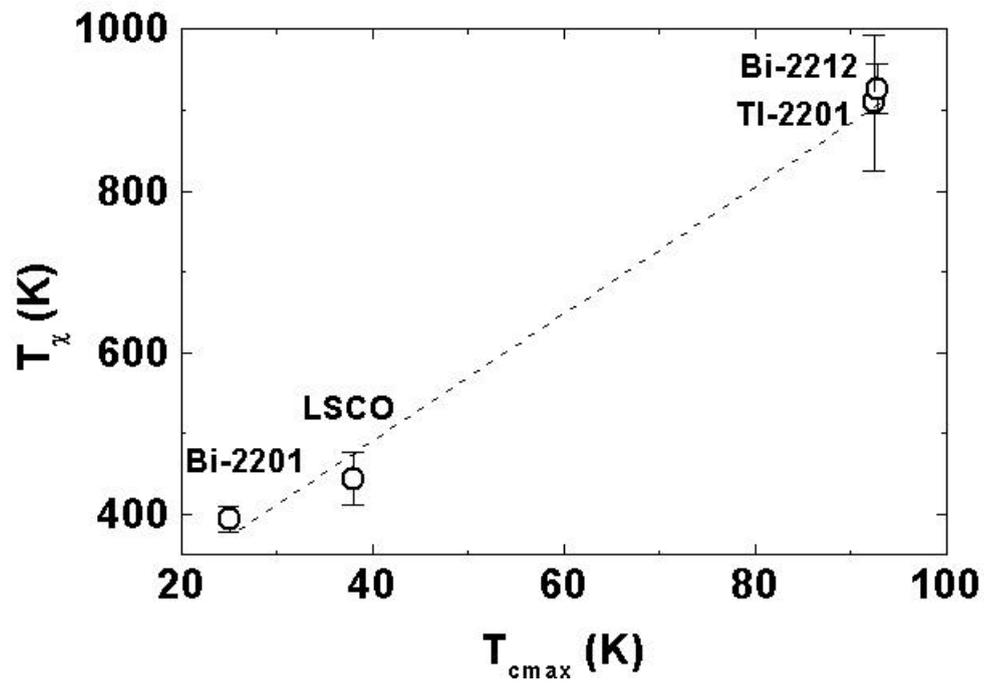

**Figure 4:** $T_c$ versus $T_{cmax}$ for the overdoped cuprates: Bi-2201(our data), LSCO(Loram), Tl-2201(Kubo) and Bi-2212 (Matsuda). The short dot line is only a guide for the eye.